\documentstyle[epsfig,longtable]{aipproc}

\begin{document}

\title{A varying kHz peak separation in 4U~1608--52}
 
\author{M.~M\'endez$^{1,2}$,
        M.~van~der~Klis$^{1}$,
        J.~van~Paradijs$^{1,3}$,
        W.~H.~G.~Lewin$^{4}$,
        B.~A.~Vaughan$^{5}$,
        E.~Kuulkers$^{6}$,
        W.~Zhang$^{7}$,
        F.~K.~Lamb$^{8}$,
        D.~Psaltis$^{9}$}

\address{
$^{1}$Astronomical Institute ``Anton Pannekoek'', University of
       Amsterdam and Center for High-Energy Astrophysics,
       Kruislaan 403, NL-1098 SJ Amsterdam, the Netherlands\\
$^{2}$Facultad de Ciencias Astron\'omicas y Geof\'{\i}sicas,
       Universidad Nacional de La Plata, Paseo del Bosque S/N,
       1900 La Plata, Argentina\\
$^{3}$Physics Department, University of Alabama in Huntsville,
       Huntsville, AL 35899, USA\\
$^{4}$Massachusetts Institute of Technology, Center for Space
       Research, Room 37-627, Cambridge, MA 02139, USA\\
$^{5}$Space Radiation Laboratory, California Institute of
       Technology, MC 220-47, Pasadena CA 91125, USA\\
$^{6}$Astrophysics, University of Oxford, Nuclear and
       Astrophysics Laboratory, Keble Road, Oxford OX1 3RH, United
       Kingdom\\
$^{7}$Laboratory for High Energy Astrophysics, Goddard
       Space Flight Center, Greenbelt, MD 20771, USA\\
$^{8}$Departments of Physics and Astronomy, University of
       Illinois at Urbana-Champaign, Urbana, IL 61801, USA\\
$^{9}$Harvard-Smithsonian Center for Astrophysics,
       60 Garden Street, Cambridge, MA 02138, USA
}

\maketitle

\begin{abstract}

Using a new technique to improve the sensitivity to weak Quasi-Periodic
Oscillations (QPO) we discovered a new QPO peak at about 1100~Hz in the
March 1996 outburst observations of 4U~1608--52, simultaneous with the
$\sim 600 - 900$~Hz peak previously reported from these data.  The
frequency separation between the upper and the lower QPO peak varied
significantly from $232.7 \pm 11.5$~Hz on March 3, to $293.1 \pm 6.6$~Hz
on March 6.  This is the first case of a variable kHz peak separation in
an atoll source.

\end{abstract}

\section*{Introduction}

Observations with the Rossi X-ray Timing Explorer (RXTE) have so far led
to the discovery of kilohertz quasi-periodic oscillations (kHz QPOs) in
about a dozen low-mass X-ray binaries (LMXB; see van der Klis 1997
\cite{vanderklis97a} for a review).  In most cases the power spectra of
these sources show twin kHz peaks that move up and down in frequency
together, keeping a constant separation (e.g., Wijnands et al.  1997
\cite{wijnandsetal97}).  In several sources a third peak has been found
near a frequency equal to the separation of the twin peaks, or twice
that value, suggesting a beat-frequency interpretation for the kHz QPOs
\cite{strohmayer96b}.

Here we present the results of the analysis of two RXTE observations of
the atoll source 4U~1608--52 on 1996 March 3 and 6, during an outburst.
Using a new technique to increase the sensitivity to weak QPOs, we
detected a second peak at $\sim 1100~$Hz, simultaneous with the $\sim
600 - 900$~Hz peak discovered by Berger et al.  (1996) \cite{berger96}
in the same data.

\section*{Detection of the second QPO}

We divided the high-time resolution data in segments of 64~sec and
calculated a power spectrum for each segment.  During both observations
the strong peak previously reported by Berger et al.  (1996)
\cite{berger96} was well detected in each segment.  Its frequency varied
from $\sim 820$ to $\sim 890$~Hz on March 3, and between $\sim 650$ and
$\sim 870$~Hz on March 6.  We fitted the centroid frequency of the
strong peak in the power spectrum of each segment, and we then shifted
the frequency scale of each individual spectrum to a frame of reference
where the position of the strong peak was constant in time.  Finally, we
averaged these shifted power spectra.

\begin{figure}[hb] 
\centerline{\psfig{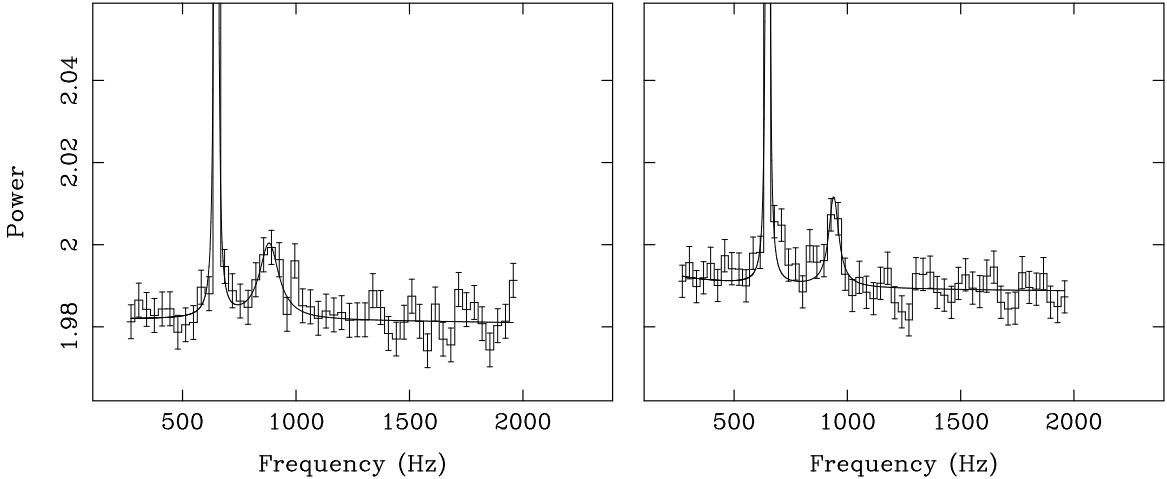}}
\vspace{10pt}
\caption{
Power spectra for the observations of 1996 March 3 (left panel) and
March 6 (right panel). The frequency of the
strong peak was arbitrarily shifted to be the same in both power
spectra
(see text for details).  On March 3 the data
cover the full PCA energy band (2 -- 60 keV), while on March 6 only
data from 2 to 12.7~keV were available.
\label{psfig}}
\end{figure}
 
If the frequency separation between the two peaks were constant, then
this ``shift and average'' procedure to compensate for the frequency
change of the strong peak would also compensate for the frequency change
of the weak peak, optimizing chances to detect it.  The improvement in
the sensitivity comes from the fact that the signal-to-noise ratio $S/N$
of a QPO peak of given rms amplitude is inversely proportional to the
square root of its width \cite{vanderklis89}, and the motion of the
peak, if uncorrected, makes it much wider, reducing $S/N$.

In Fig.~\ref{psfig} we show both power spectra calculated using the
above method.  In both cases a second QPO peak can be seen at a
frequency $\sim 200 - 300$~Hz above that of the strong peak of Berger et
al.  (1996) \cite{berger96}.  We fitted each power spectrum with a
function consisting of a constant level, representing the Poisson noise,
the Very Large Event contribution \cite{zhang95i,zhang95ii}, and two
Lorentzians.  The $1\sigma$ error bars from the fits indicate the second
peak to be $4.3 \sigma$ and $4.4 \sigma$ significant on March 3, and
March 6, respectively.  An $F$-test to the $\chi^{2}$ of the fits with
and without this peak yields a probability of $7.1 \times 10^{-10}$ on
March 3, and $5.3 \times 10^{-8}$ on March 6, for the null hypothesis
that the peak is not present in the data.  Considering the number of
trials implied by the number of independent frequencies analyzed
\cite{vanderklis89}, these probabilities increase to $1.4 \times
10^{-7}$ and $1.1 \times 10^{-5}$, respectively.

Interestingly, $\Delta \nu$ changed from $232.7 \pm 11.5$~Hz on March 3
to $293.1 \pm 6.6$~Hz on March 6, a change of $60.4 \pm 13.3$~Hz.  We
tested the significance of this result by fitting both power spectra
simultaneously, but forcing the distance between the peaks to be the
same in both of them.  Applying an $F$-test to the $\chi^{2}$ of this
fit and the fit where all parameters were free we get a probability of
$2.4 \times 10^{-3}$ for the hypothesis that the peak separation did not
change between the two observations:  the difference in the frequency
separation between March 3 and 6 is significant at the $3.2\sigma$
level.

\begin{figure}[hb] 
\centerline{\psfig{file=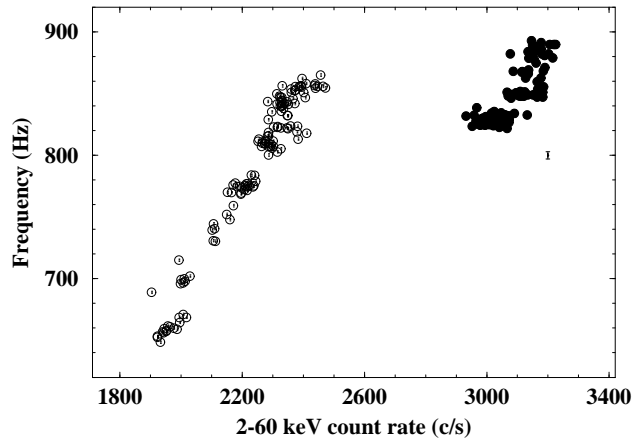,width=3.0in,angle=-90}}
\vspace{10pt}
\caption{
The QPO frequency vs.  count rate.  The data for March 3 and 6 are
plotted using filled and open circles, respectively.
\label{freq_rate}}
\end{figure}

In Fig.~\ref{freq_rate} we plot QPO frequency versus count rate for 1996
March 3, 6.  In both cases frequency is positively correlated to the
source count rate, however there is no simple function that fits both
observations simultaneously.

A similar effect has been observed when different sources, spanning a
very large range of luminosities, are plotted together in a single
frequency-luminosity diagram:  each source shows a positive correlation,
along lines which are more or less parallel (e.g., van der Klis et al.
1997a \cite{vanderklisetal97a}).  A similar behavior as we see in
4U~1608--52 was also observed in 4U~0614+091 \cite{ford97a,mendez97a}.
The fact that this is observed in individual sources shows that a
difference in neutron star properties such as mass or magnetic field
strength can not be the full explanation for the differences observed in
the frequency-luminosity relations.

\section*{Conclusions}

We have found, for the first time, the second peak in 4U~1608--52
expected on the basis of comparison to other kHz QPO sources.  We see
the separation between the two peaks vary.  In the case of Sco~X--1,
where $\Delta \nu$ also varies \cite{vanderklisetal97b}, it has been
argued that the variations can be attributed to near-Eddington
accretion.  This explanation can not apply to 4U~1608--52, as its
luminosity was $1.3 \times 10^{37}$~erg~s$^{-1}$ and $9.4 \times
10^{36}$~erg~s$^{-1}$ on March 3 and March 6, respectively
\cite{mendez97b}, less than 10\,\% of $L_{\rm Edd}$.  The dependence of
the QPO frequency on count rate is complex.  Although for each
individual observation frequency and count rates are correlated, no
simple relation fits both observations simultaneously.  The frequency of
the QPO did not change much while the source count rate dropped by
20\,\%.

\end{document}